\documentstyle[11pt,aaspp4,psfig]{article}

\def\dash{\hbox{--}}
\def\degree{{$^\circ$}}

\def\puncspace{\ifmmode\,\else{\ifcat.\C{\if.\C\else%
\if,\C\else\if?\C\else\if:\C\else\if;\C\else\if-\C\else%
\if)\C\else\if/\C\else\if]\C\else\if'\C%
\else\space\fi\fi\fi\fi\fi\fi\fi\fi\fi\fi}%
\else\if\empty\C\else\if\space\C\else\space\fi\fi\fi}\fi}%
\def\SP{\let\\=\empty\futurelet\C\puncspace}

\def\ee#1{\ifmmode {} \times 10^{#1} \else ${} \times 10^{#1}$\fi}
\def\sub#1{\ifmmode _{#1} \else $_{#1}$\fi}
\def\sup#1{\ifmmode ^{#1} \else $^{#1}$\fi}

\def\dash{\hbox{--}}
\def\about{\ifmmode \sim \else {$\sim\,$}\fi}
\def\lta{\ifmmode {\,\mathbin{\lower 3pt\hbox   
    {$\,\rlap{\raise 5pt\hbox{$\char'074$}}\mathchar"7218\,$}}}
    \else {${\mathbin{\lower 3pt\hbox
    {$\rlap{\raise 5pt\hbox{$\char'074$}}\mathchar"7218\,$}}}
    $}\fi}
\def\gta{\ifmmode {\mathbin{\lower 3pt\hbox   
    {$\,\rlap{\raise 5pt\hbox{$\char'076$}}\mathchar"7218\,$}}}
    \else {${\mathbin{\lower 3pt\hbox
    {$\rlap{\raise 5pt\hbox{$\char'076$}}\mathchar"7218\,$}}}
    $}\fi}

 \mathcode`*="002A   

\def\degree{{\ifmmode ^\circ \else $^\circ$\fi}}

\def\mdot{{\ifmmode \dot M \else {$\dot M$}\fi}}
\def\mdote{{\ifmmode \dot M_E \else {$\dot M_E$}\fi}}
\def\mdoti{{\ifmmode \dot M_i \else {$\dot M_i$}\fi}}
\def\msun{{\ifmmode M_\odot \else {$M_{\odot}$}\fi}}

\begin{document}

\lefthead{Miller}
\righthead{Effects of Radiation Forces on Precession}

\title{Effects of Radiation Forces on the Frequency of\\
Gravitomagnetic Precession Near Neutron Stars}

\author{M.\ Coleman Miller}
\affil{Department of Astronomy and Astrophysics, University of Chicago\\
       5640 South Ellis Avenue, Chicago, IL 60637, USA\\
       miller@bayes.uchicago.edu}
\authoremail{miller@bayes.uchicago.edu}

\begin{abstract}

Gravitomagnetic precession near neutron stars and black holes has received
much recent attention, particularly as a possible explanation of
15--60~Hz quasi-periodic brightness oscillations (QPOs) from
accreting neutron stars in low-mass X-ray binaries, and of somewhat
higher-frequency QPOs from accreting stellar-mass black holes. Previous
analyses of this phenomenon have either ignored radiation
forces or assumed for simplicity that 
the radiation field is isotropic, and in particular that there is 
no variation of the radiation field with angular distance from the 
rotational equatorial plane of the compact object. However,
in most realistic accretion geometries (e.g., those in which the
accretion proceeds via a geometrically thin disk) the radiation
field depends on latitude. Here we show that in this case
radiation forces typically have an important, even dominant,
effect on the precession frequency of test particles in orbits 
that are tilted with respect to the star's rotational equator.
Indeed, we find that even for accretion luminosities only a few percent
of the Eddington critical luminosity, the precession
frequency near a neutron star can be changed by factors of up to
$\sim 10$. Radiation forces must therefore be included in analyses
of precession frequencies near compact objects, in such varied 
contexts as low-frequency QPOs, warp modes of disks, and trapped
oscillation modes.  We discuss specifically the impact
of radiation forces on models of low-frequency QPOs involving
gravitomagnetic precession, and show that such models are rendered 
much less plausible by the effects of radiation forces.

\end{abstract}


\section{INTRODUCTION}

The microsecond time resolution, $\sim$6000~cm$^2$ effective area,
and $\sim$256~kbps telemetry capability of the {\it Rossi} X-ray
Timing Explorer (RXTE) have made possible the discovery of
brightness oscillations with frequencies $\sim$300--1200~Hz in
both accretion-powered and thermonuclear-powered emission 
(see, e.g., van der Klis 1997 for a review of the properties
of these oscillations) from
some sixteen neutron-star low-mass X-ray binaries (LMXBs). The
remarkably coherent brightness oscillations observed during
type~I (thermonuclear) X-ray bursts are thought to occur at the
stellar spin frequency or its first overtone (see, e.g., 
Strohmayer, Zhang, \& Swank 1997). The commonly observed
pairs of quasi-periodic brightness oscillations (QPOs) in
the accretion-powered emission are generally thought to be
generated by a beat-frequency mechanism, in which the 
frequency of the higher-frequency QPO peak in a pair is
the Keplerian orbital frequency at a special radius near the
neutron star, and the frequency of the lower-frequency peak is 
the difference between this Keplerian frequency and the stellar
spin frequency (Miller, Lamb, \& Psaltis 1998; Strohmayer et
al.\ 1996). The
observations and modeling of these brightness oscillations have
produced a rapid advance in our understanding of these 
systems. For example, if these interpretations of the brightness
oscillations are correct then we know, for the first time, 
the spin frequencies of
more than a dozen neutron stars in LMXBs, and as a result the
evolutionary connection between LMXBs and millisecond pulsars
has been strengthened greatly. Even more dramatic is the 
likelihood that the properties of the brightness oscillations provide
robust and important constraints on the equation of state of
neutron star matter (see, e.g., Miller et al.\ 1998).

The flood of new information from RXTE has also led to a re-examination
of the physical picture of neutron-star LMXBs that was developed
earlier, based on the 2--20~keV energy spectra and 1--100~Hz
power spectra of these sources obtained using
satellites such as EXOSAT and Ginga. A phenomenon that has played
an especially useful role in the development of this picture is
the so-called horizontal branch oscillations, or HBOs (van der 
Klis et al.\ 1985; see van der Klis 1989 for a review), which are 
a type of QPO that is observed in the horizontal branch
spectral state of the persistently brightest neutron-star LMXBs 
(the Z sources). These oscillations have frequencies 
$\nu_{\rm HBO}\sim$15--60~Hz, fractional rms amplitudes of a few
percent, and coherences $\nu_{\rm HBO}/\Delta\nu_{\rm HBO}
\sim$2--10, where $\Delta\nu_{\rm HBO}$ is the FWHM of the peak in
the power spectrum. The most successful model of HBOs is the
magnetospheric beat frequency model (Alpar \& Shaham 1985; 
Lamb et al. 1985; Shibazaki \& Lamb 1987), in which the 
observed frequency is the
difference between the stellar spin frequency and the orbital
frequency at the radius in the accretion disk at which the stellar
magnetic field picks up and channels gas from the disk onto the
magnetic polar regions. This model accounts for many of the main
features of HBOs, including their range of frequencies, their
amplitudes, and their dependence on inferred mass accretion rate.
Moreover, the stellar magnetic moments that are required in this
model are confirmed independently by model fits to the 
energy spectrum (Psaltis, Lamb, \& Miller 1995; Psaltis \& Lamb 1998),
and the predicted stellar spin frequencies
(Ghosh \& Lamb 1992) are consistent with the $\sim 300$~Hz spin
frequencies inferred from observations.

However, observations using RXTE and further theoretical modeling
have turned up aspects of the HBOs that have raised questions
about the magnetospheric beat-frequency
model. For example, the FWHM of HBOs is often
10~Hz or less (see van der Klis 1989), which is only a fraction 
$\sim 0.03$ of the stellar spin frequency $\nu_{\rm spin}\sim$300~Hz. 
Early analytical work (Ghosh \& Lamb 1978, 1979) suggested that the 
fractional width of the magnetospheric transition region could indeed 
be of order $0.02\dash 0.05$. More recent numerical modeling, however,
has yielded much larger estimates for the fractional width,
$\Delta r/r\sim$0.1--0.3 (Daumerie \& Lamb 1998). 
In addition, Stella \& Vietri (1998) have pointed out that, over
a limited range of frequencies, the
Z sources GX~17$+$2 and GX~5$-$1 display an interesting correlation between
their HBO frequency and the frequency of their higher-frequency QPO
peak in a pair: $\nu_{\rm HBO}\propto\nu_{\rm kHz}^2$.
Such a correlation, if it indeed represents a physically important
connection between the HBOs and the kilohertz QPOs,
is not explained a priori in a magnetospheric beat frequency
model.

Stella \& Vietri (1998; also 
Stella 1997a,b) have suggested an entirely new explanation for the 
HBOs. In their interpretation, the frequency is the gravitomagnetic,
or Lense-Thirring, precession frequency, which is the frequency
at which an orbit tilted with respect to the stellar spin axis
will precess about the spin axis. In this interpretation,
$\nu_{\rm HBO}\propto\nu_{\rm kHz}^2$ is predicted if the radius at
which this frequency is generated is the same radius at which
the kilohertz QPOs are generated, although the proportionality 
constant derived from fits of this formula to observations is
roughly two to four times (depending on the symmetry) the constant
expected for the most realistic equations of state. The possible
match of the observed frequency behavior is intriguing, but as
yet no convincing physical mechanism has been suggested that
would generate QPOs at the Lense-Thirring precession frequency.

To have precession at all requires
that the gas generating the QPO be in an orbit that is tilted
with respect to the spin
equator of the star. One may then categorize possible mechanisms
by whether they involve gas that is coupled over a range of
radii (e.g., as a warp in a
disk) or not coupled (e.g., as a thin annulus or orbiting clumps
of gas decoupled from each other). Markovic \& Lamb (1998) have shown that
disk modes extending over a range of radii cannot generate
the observed QPOs, for two reasons. First, the frequencies of
the modes (1~Hz at most) are too low to explain the observed
QPOs. This is true even of modes that are driven by, e.g., 
magnetic or radiative stresses. Second, the modes are
very heavily damped, with damping rates thousands of times
greater than the precession frequency. This still leaves open
the possibility of precessing gas that is decoupled except over
a small range of radii.

Here we
calculate the precession frequency of a test particle in an inclined 
orbit around a rotating and radiating neutron star. If the star does
not radiate, then the precession frequency is just the Lense-Thirring
frequency. However, these stars {\it are} radiating; indeed, the
Z sources have luminosities that are inferred to be at least tens of
percent of the Eddington critical luminosity when the HBO is
observed. Radiation forces can therefore be extremely important.
Here we calculate their effect on the precession frequency.
We show that radiation forces can change the precession frequency 
dramatically, by factors of several, even if the luminosity is
only a few percent of the Eddington luminosity. This is especially
true if, as expected, the radiation field is not independent of
latitude (e.g., if the accretion occurs in an equatorial band
instead of uniformly over the entire surface of the neutron star).
Also, even relatively
small azimuthal variations in the brightness at a given
radius can reduce sharply the coherence of any resulting QPO.
Hence, to explain the observed coherence of the HBOs in a model
involving Lense-Thirring precession, one must explain why the 
radiation field is azimuthally symmetric to high accuracy. The 
axisymmetry requirements are not so severe for a Keplerian-frequency
brightness oscillation, such as is thought to generate the
higher-frequency QPO observed in a pair, because the
orbital frequency is not nearly so sensitive to radiation
forces.

In \S~2 we develop the methods used to estimate the effects of
radiation forces on precession frequencies. We give our results
and discuss their implications for HBOs in \S~3.

\section{METHOD}

Throughout this paper, we use the procedures and notation given
in Miller \& Lamb (1996; see also Abramowicz, Ellis, \& Lanza
1990). In particular, we assume that
the only interaction between radiation and the test particle
occurs via isotropic, frequency-independent scattering, which is
a good approximation for the frequencies and ionization
fractions expected near accreting neutron stars (see Lamb \& Miller
1995 for further discussion). Hence, the radiation force is

\begin{equation}
f^\alpha=\sigma F^\alpha\; ,
\end{equation}
where $\sigma$ is the scattering cross section,
\begin{equation}
F^\alpha=-T^{\mu\alpha}u_\mu-u^\alpha T^{\mu\beta}u_\mu u_\beta
\end{equation}
is the radiative energy flux measured in the rest frame of the
particle (see Miller \& Lamb 1996), and $T^{\mu\alpha}$ are the
components of the radiation stress-energy tensor at a given
event in the Boyer-Lindquist coordinate system. Here and below
we set $G=c\equiv 1$ except where noted.

Let us assume that a particle moving in an almost circular
orbit near the spin equatorial plane, is perturbed
by a small vertical velocity, which is therefore in the 
$\theta$-direction. The four-velocity is then

\begin{equation}
u^\mu=(u^t,0,u^\theta,u^\phi)\; ,
\end{equation}
where $u^t$ and $u^\phi$ are the same as for a circular equatorial
orbit (note, however, that $u^t$ and $u^\phi$ are modified by
radial radiation forces, and are therefore different from their
values in the absence of radiation). The $\theta$-component of
the equation of motion is then 

\begin{equation}
{d^2\theta\over{d\tau^2}}+{1\over 2}g^{\theta\theta}
(g_{\theta\mu,\nu}+g_{\theta\nu,\mu}-g_{\mu\nu,\theta})
u^\mu u^\nu=f^\theta/m\; .
\end{equation}

We expand only to first order in the dimensionless spin
parameter $j\equiv cJ/GM^2$, because to higher order in $j$
the spacetime external to the star must be calculated
numerically (in particular, to order $j^2$ and higher the
spacetime deviates from the Kerr spacetime; see, e.g., Cook,
Shapiro, \& Teukolsky 1994). To this order, and assuming that
the angular deviation $\delta$ from the equatorial plane
is small (i.e., $\delta\ll 1$), we find

\begin{equation}
{d^2\theta\over{d\tau^2}}=\left[\left(u^\phi\right)^2
-{4jM^2\over{r^3}}u^t u^\phi\right]\delta-{\sigma\over m}
\left[T^{\mu\theta}u_\mu+u^\theta T^{\mu\beta}u_\mu u_\beta\right]\; .
\end{equation}
Dividing through by $(u^t)^2$ and using the fact that in 
Boyer-Lindquist coordinates $u^\phi/u^t=\Omega$, the angular velocity
as observed at infinity, we find after some further manipulation
(see also Kato 1990)

\begin{equation}
{d^2\theta\over{dt^2}}=-\Omega_\perp^2\delta
-{\sigma\over m}\left(u^t\right)^{-2}\left[
T^{\mu\theta}u_\mu+u^\theta T^{\mu\beta}u_\mu u_\beta\right]\; ,
\end{equation}
where $v^\theta\equiv u^\theta/u^t$, $\Omega_K$ is the angular
velocity of a circular orbit as measured at infinity, and

\begin{equation}
\Omega_\perp^2=\Omega_K^2-{4jM^2\over{r^3}}\Omega_K
\end{equation}
is the vertical epicyclic frequency in the absence of radiation
forces, to first order in $j$. The Lense-Thirring precession 
frequency is simply the difference between the vertical epicyclic
frequency and the orbital frequency $\Omega_K$.

In the $\theta$-component of the force equation,
the radiation terms may be divided into the velocity-independent
flux term $T^{t\theta}u_t$ and the remaining terms, which are
velocity-dependent. The velocity-dependent terms are ``drag"
terms and are dissipative. That is, for example, the amplitude of 
vertical oscillatory motion about the equatorial plane is
damped by these terms. In contrast, the velocity-independent
term is non-dissipative and changes only the frequency, not
the amplitude, of the vertical oscillatory motion. If the
radiation from the star is assumed to be isotropic,
as is a standard assumption in
many treatments of warped disk modes near compact objects
(see, e.g., Pringle 1996; Markovic \& Lamb 1998), then there
is no net flux in the $\theta$-direction and $T^{t\theta}=0$.
In general, however, the radiation field will not be isotropic,
and will have a $\theta$-dependence and hence a net flux in
the $\theta$-direction. For example, if accretion occurs via
a thin disk then the radiation intensity near the rotation
equator is greater than the radiation intensity far from the
equator, and there is therefore a gradient of flux away from
the equatorial plane.

In such a case, it is typical that for small $\theta$-displacements
and hence small $u^\theta$ that the velocity-independent term

\begin{equation}
T^{t\theta}u_t=(1-2M/r)^{-1/2} r^{-1}T^{\hat t\hat\theta}u_t
\end{equation}
is the largest of the radiation terms by two or more orders of
magnitude. It is therefore likely that, for a realistic radiation
pattern, the frequency (but not the amplitude) of vertical
epicyclic motion will be changed significantly by radiation
forces. As we show below, for displacements $\delta\ll 1$
the radiation force is proportional to $\delta$, call it
$\kappa_{\rm rad}\,\delta$, and hence the vertical
epicyclic frequency becomes, schematically,

\begin{equation}
\Omega_{\perp,{\rm rad}}^2=
\Omega_K^2-{4jM^2\over{r^3}}\Omega_K-\kappa_{\rm rad}\; .
\end{equation}
We compute $\kappa_{\rm rad}$ in a simplified model, for one 
particular geometry, in \S~3.1 (equation [\ref{radforce}]).
Note that even if the epicyclic frequency is modified only slightly
by the radiation forces, the precession frequency (which is
$\Omega_K-\Omega_{\perp,{\rm rad}}$) can be changed dramatically. 
Indeed, as we now show, radiation forces can change the precession 
frequency by factors of several, even for luminosities that are
only a few percent of the Eddington critical luminosity.
In the next section we estimate the effects of radiation
forces for one specific pattern of emission on the stellar surface,
to give an idea of the magnitude and dependences. We then discuss
the implications that these effects have for models of HBOs
involving Lense-Thirring precession.

\section{RESULTS AND DISCUSSION}

\subsection{Magnitude of Radiation Forces in Simplified Model}

Consider a nonrotating star that radiates uniformly from a band
of angular half-width $\epsilon$ around the equator, and does not radiate
from any other part of the surface. Let the specific intensity $I_s$
at the surface be isotropic in the outward direction. What is
$T^{\hat t\hat\theta}$, the $\theta$-component of the flux
as measured in a local tetrad?

Formally, this could be computed in a Schwarzschild spacetime
from

\begin{equation}
T^{\hat t\hat\theta}=I_s{(1-2M/R)^2\over{(1-2M/r)^2}}
\int_0^{2\pi}\int_0^{\alpha(\tilde b)}\sin^2\tilde a\,
\cos\tilde b\,d\tilde a\,d\tilde b
\label{Tschwarz}
\end{equation}
(see, e.g., Miller \& Lamb 1996), where $\alpha(\tilde b)$ is the
angular extent of the band in direction $\tilde b$ measured by
an observer at Boyer-Lindquist radius $r$. 
For an observer in the equatorial plane ($\theta=\pi/2$),
$T^{\hat t\hat\theta}=0$ because the contributions above and
below the plane exactly cancel. For an observer out of the
equatorial plane the contributions do not exactly cancel, and
the integral must be performed. Unfortunately,
in a Schwarzschild spacetime the computation of $\alpha(\tilde b)$
requires numerical integration, and does not yield much insight.
We therefore simplify further to straight-line photon propagation.
A straightforward but tedious calculation using spherical
triangles then shows that $\alpha(\tilde b)$ is given implicitly
by

\begin{equation}
\sin^2\alpha(\tilde b)=\left(R\over r\right)^2{\sin^2\psi(\tilde b)
\over{1-2(R/r)\cos\psi(\tilde b)+R^2/r^2}}\; ,
\end{equation}
where $\psi(\tilde b)$, the angle spanned by the emitting band in direction
$\tilde b$ as measured from the center of the star, is given by

\begin{equation}
\sin\psi(\tilde b)={\sin\epsilon\over{\left(1-\sin^2\tilde b\cos^2\epsilon
\right)^{1/2} }}\; .
\end{equation}

Suppose now that $\epsilon$ is small enough that we can ignore the
small angular extent (near $\tilde b=\pm\pi/2$) where
the band, in the direction $\tilde b$, extends beyond the visible
horizon. This is a good approximation for $\epsilon\lta 0.2$.
The integral (\ref{Tschwarz}) then simplifies to

\begin{equation}
T^{\hat t\hat\theta}\approx I_s{(1-2M/R)^2\over{(1-2M/r)^2}}
\int_{-\pi/2}^{\pi/2}\cos\tilde b\,d\tilde b\,
2\delta{d\alpha(\tilde b)\over{d\theta}}\sin^2\alpha(\tilde b)\; .
\label{eqthet}
\end{equation}
In equation (\ref{eqthet}) we have subtracted the contribution of
the radiation field at latitudes greater than that of the observer
($\tilde b=-\pi/2$ to $\pi/2$) from the contribution of the
radiation field at latitudes less than that of the observer
($\tilde b=\pi/2$ to $3\pi/2$). Combining these equations, we find

\begin{equation}
T^{\hat t\hat\theta}\approx \delta I_s 
{4(1-2M/R)^2\over{(1-2M/r)^2}} {R^3\over{r^3}}\cos\epsilon\sin^2\epsilon
\int_0^{\pi/2}{\cos^3\tilde b\,d\tilde b\over{
\left(1-2{R\over r}\cos\psi(\tilde b)+{R^2\over{r^2}}\right)^2
\left(1-\sin^2\tilde b\cos^2\epsilon\right)^{5/2}}}\; .
\end{equation}
This estimate is in excellent agreement with the numerical
results described in \S~3.

At large radii, we can obtain a simple analytic expression, which,
however, underestimates the radiation force at small radii
(that is, where $r\lta$few$\times R$). In the limit $r\gg R$,

\begin{equation}
T^{\hat t\hat\theta}\approx \delta I_s {8\over 3}
(1-2M/R)^2{R^3\over{r^3}}\cos\epsilon\sin\epsilon\; 
\end{equation}
and
\begin{equation}
\kappa_{\rm rad}\approx \left(I_s\over I_E\right){M\over r^3}
(1-2M/R){R\over r}\cos\epsilon\sin\epsilon\; ,
\label{radforce}
\end{equation}
where $I_E$ is the critical specific intensity
(if the star were emitting from its entire
surface with a specific intensity $I_E$, then the luminosity
measured at infinity would be the Eddington luminosity; see
also Miller \& Lamb 1996).
The ratio of the radiation term to the Lense-Thirring term
is then

\begin{equation}
{\kappa_{\rm rad}\over{4jM^2\Omega_k/r^3}}\approx
{(I/I_E)(1-2M/R)(R/M)\cos\epsilon\sin\epsilon\over{4j(M/r)^{1/2}}}\; .
\label{ratios}
\end{equation}
At $r=12\,M$ for a star with $j=0.2$, radius $5\,M$, and an
emitting band of half-width
$\sin\epsilon=0.2$, equation (\ref{ratios}) predicts that the 
two terms are equal when $I_s=0.4\,I_E$. In fact, the effects 
of radiation 
increase very rapidly with decreasing radius, and the two terms
are instead approximately equal when $I_s=0.1\,I_E$. Note that,
because we have assumed an emitting band with an angular 
half-width $\sin\epsilon=0.2$, this intensity corresponds to a
luminosity of only $L=0.02\,L_E$.

This analysis shows that radiation forces can have
a major impact on the precession frequency near neutron stars.
We now give numerical results for this precession
frequency, as a function of radius and radiation intensity.

\subsection{Numerical Results}

A detailed description of the codes used to calculate the radiation
stress-energy tensor and follow the motion of test particles around
rotating, radiating neutron stars is given in Miller \& Lamb (1996).
In essence, the codes simply calculate the stress-energy tensor by
ray tracing, then calculate the motion of test particles using the
relativistic force equation. In the calculations reported here, we
use a radiating band, centered on the rotational equator, of 
half-width 0.2 times the radius of the star. The stellar radius is
set to $R=5\,M$, and the rotation parameter is $j=0.2$. This is
a high value of $j$; for example, a $1.8\,M_\odot$ star with a
spin frequency $300$~Hz has $j\approx 0.1$ for realistic equations
of state. Therefore, if radiation forces alter the precession
frequency significantly for the extreme value of $j=0.2$, the 
effects will be even more important for smaller rotation parameters.
For the external spacetime we use the Kerr spacetime, which for $j=0.2$
is insignificantly different from the true spacetime around a
rotating neutron star. These calculations therefore neglect the
corrections caused by classical precession, which are small for
the radii and spin rates of interest (see Stella \& Vietri 1998).

Figure~1 shows the results of these calculations. In this figure,
we plot the precession frequency as seen at infinity as a function
of the luminosity at infinity, $L/L_E=0.2\,I_s/I_E$. For comparison,
we also plot the analytic value of the Lense-Thirring frequency for
zero radiation (solid line). In Figure~2 we focus on the radial
dependence of the frequency for two different luminosities, with
the $r^{-3}$ dependence of the Lense-Thirring frequency divided
out. Clearly, close to the star the radiation component of the
precession frequency increases much faster than does the 
gravitomagnetic component. In Figure~3 we focus on the effect
of increasing the luminosity, at a fixed radius $r=10\,M$
and with $R=5\,M$, $j=0.2$, and a radiating band with fractional
half-width 0.2, as before.

\subsection{Discussion}

These figures demonstrate that radiation forces can produce a
precession frequency that is many times the gravitomagnetic
precession frequency. Here we have presented results for one
particular emission pattern, and for test particles. However, the main
result of this paper, that precession frequencies can be changed
dramatically by radiation forces, is much more generally applicable.
For example, similar qualitative effects (although differing in
detail) are to be expected when radiation is
absorbed by the accretion disk and then reradiated outward from
the disk. There are, therefore, circumstances in which the vertical
component of the radiation force could alter the stability or
properties of radiation-driven warps around stars. For instance,
if the angle subtended by the radiating layer on the star, as
seen at some radius $r$, is comparable to the warp angle of the
disk, then the radiation absorbed by the disk is increased by
a factor of a few. The torque on the disk would therefore be
change by similar factors. Hence, close to radiating stars the
radiation-driven warping of disks can be altered by vertical
radiation forces (for discussion of the importance
of warping in a variety of astrophysical situations see, e.g., 
Pringle 1996, 1997; 
Maloney, Begelman, \& Pringle 1996; Livio \& Pringle 1997; Armitage
\& Pringle 1997; Maloney \& Begelman 1997; Markovic \& Lamb 1998).

The test particle approximation is strictly valid only 
if the optical depth to the stellar surface is much less than
unity. If instead there is significant shadowing in some directions,
the qualitative effects of radiation forces can be different then
they are in the simple calculations presented here. For example,
if there is substantial emission above and below the disk but
shadowing decreases emission in the midplane, the epicyclic frequency
is increased by radiation forces, and hence the precession frequency
is decreased. What can be said generally is that, because
the magnitude of the radiation term can exceed the magnitude of
the gravitomagnetic term even for low luminosities, the precession
frequency is highly sensitive to details of the radiation field.

This has important consequences for any model of HBOs that invokes
precession. For instance, it means that for many realistic emission
patterns the precession frequency is much greater than the
Lense-Thirring frequency. As is clear from Figure~2, the precession
frequency will therefore depend strongly on both luminosity and
radius (indeed, even at a fixed luminosity the precession frequency 
is much steeper than $r^{-3}$ near the star).
Hence, it would require an improbable
coincidence for the precession frequency to have an $r^{-3}$ 
dependence on radius, as has been claimed for GX~17$+$2 
and GX~5$-$1 (Stella \& Vietri 1998).

The production of narrow HBOs by Lense-Thirring precession also 
requires that the radiation field be very axisymmetric. 
Consider for example a Z source radiating at a few tens of percent
of the Eddington luminosity. From the results in \S~3.1 and \S~3.2 
it is clear that a fractional azimuthal variation 
in the intensity of only $\Delta I/I\sim 0.05\dash 0.1$ would
produce a much larger FWHM to the QPO than is observed,
because the change in precession frequency would exceed the
Lense-Thirring precession frequency. Such a fractional azimuthal
variation arises in many plausible scenarios; for example, this
could happen if the optical depth from that radius to the stellar
surface varies slightly with azimuth. In contrast, the orbital frequency, 
which in this picture gives the higher-frequency QPO peak in a pair,
is much less sensitive (see \S~5 of Miller et al.\ 1998), and in
such a situation would vary by less than 1\%. 

Coherence of the HBO is even more difficult to maintain in a 
Lense-Thirring precession model if the HBO is generated by the
precession of the footprint of impact on the stellar surface.
This is because, in that case, one needs to map the precession frequency
at some orbital radius onto the stellar surface. As discussed in
Miller et\ al. (1998), this demands near axisymmetry from the
entire spiral of gas from the orbital radius to the surface, and
not just from the movement of gas near the orbital radius. In
turn, this means that the radiation field must be nearly 
axisymmetric from the orbital radius inward, because otherwise
the gas from even a single clump would become dephased. This
appears difficult, particularly given that radiation stresses
have a rapidly increasing effect on the precession frequency of
gas as the gas moves closer to the star. Note that this problem
does not exist for the magnetospheric beat frequency model of
HBOs, because in that model the flow of gas onto the star is
controlled by the magnetic field, and hence radiation forces
have only a minor effect.

In conclusion, we have shown that radiation forces have an
extremely strong effect on orbital precession near accreting
neutron stars. Combined with recent results on warped disk modes
near neutron stars (Markovic \& Lamb 1998), this makes
Lense-Thirring explanations of neutron-star QPOs much less promising 
than had been thought previously. More generally, vertical radiation 
forces may modify torques or precession frequencies in other
contexts as well, and they must therefore be considered in
treatments of warps or disk modes around accreting neutron stars
and black holes.

\acknowledgements

We are grateful to Dimitrios Psaltis for many helpful discussions
of the properties of horizontal branch oscillations, and to Fred
Lamb for useful comments about an earlier draft of this paper.
This work was supported in part by NASA grant NAG~5-2868 at the
University of Chicago.

\newpage

\figcaption[]{
Precession frequencies as a function of
radius, for various luminosities. The left axis gives
the frequency in units of $M^{-1}$, and the right axis gives the
frequency in Hertz, assuming a stellar gravitational mass
$M=1.8\,M_\odot$. As discussed in the 
text, these calculations were carried out assuming a rotation parameter
$j=0.2$, a stellar radius $R=5\,M$, and a uniformly radiating band
of half-width 0.2 times the stellar radius. The solid line gives
the analytical Lense-Thirring precession frequency, calculated
exactly in a Kerr spacetime with $j=0.2$. The dotted line gives 
the numerically
computed precession frequency for no radiation, for comparison
with the analytical frequency. The long dashed line gives
the precession frequency for a luminosity 0.02 times the
Eddington critical luminosity, and the dash-dotted line gives
the frequency for a luminosity 0.04 times the Eddington
luminosity. This figure shows that the precession
frequency can be changed dramatically by radiation forces,
particularly for radii close to the star.}

\figcaption[]{
Ratio of precession frequencies to the Lense-Thirring frequency,
for the same luminosities and parameters as in Figure~1. This
figure highlights the deviation from the $r^{-3}$ scaling of
the precession frequency at small radii.}

\figcaption[]{
Precession frequencies as a function of luminosity (dashed line), 
at the fixed radius $r=10\,M$. As in Figure~1,
we assume $j=0.2$, a stellar radius $R=5\,M$, and a uniformly radiating 
band of half-width 0.2 times the stellar radius. The solid
horizontal line gives the precession frequency in the absence of
radiation forces. The left and right
hand axes are as in Figure~1, except that here the frequencies are
plotted linearly instead of logarithmically. As expected from the
analytical treatment in \S~3.1, for low luminosities the deviation
from the Lense-Thirring precession frequency is linear, but at
high luminosities the precession frequency is higher than a simple
linear extrapolation would predict.}

\begin{figure*}[t]
\hbox{\hskip -0.5truein
\psfig{file=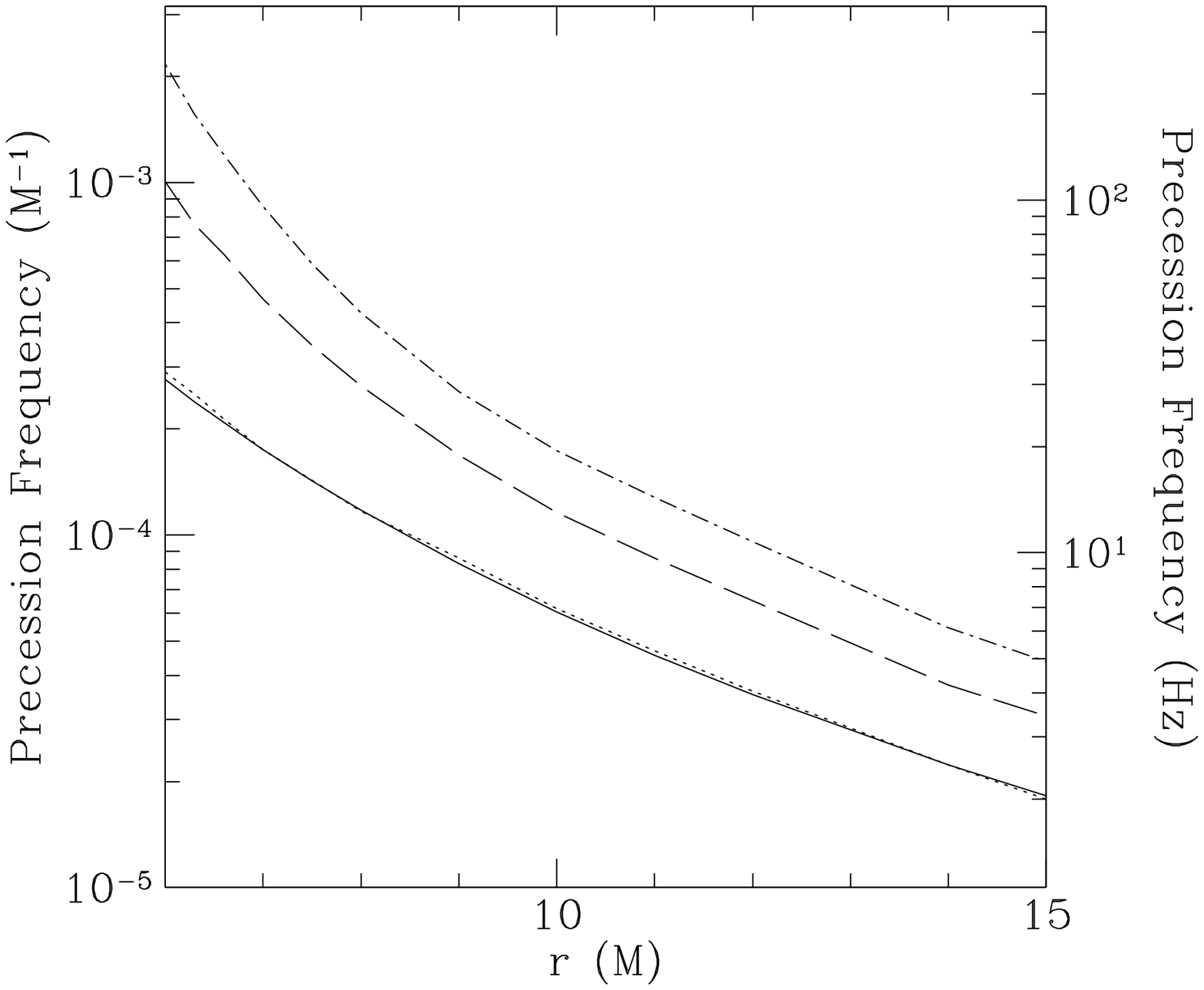,height=8.0truein,width=8.0truein}}
\end{figure*}

\begin{figure*}[t]
\hbox{\hskip -0.5truein
\psfig{file=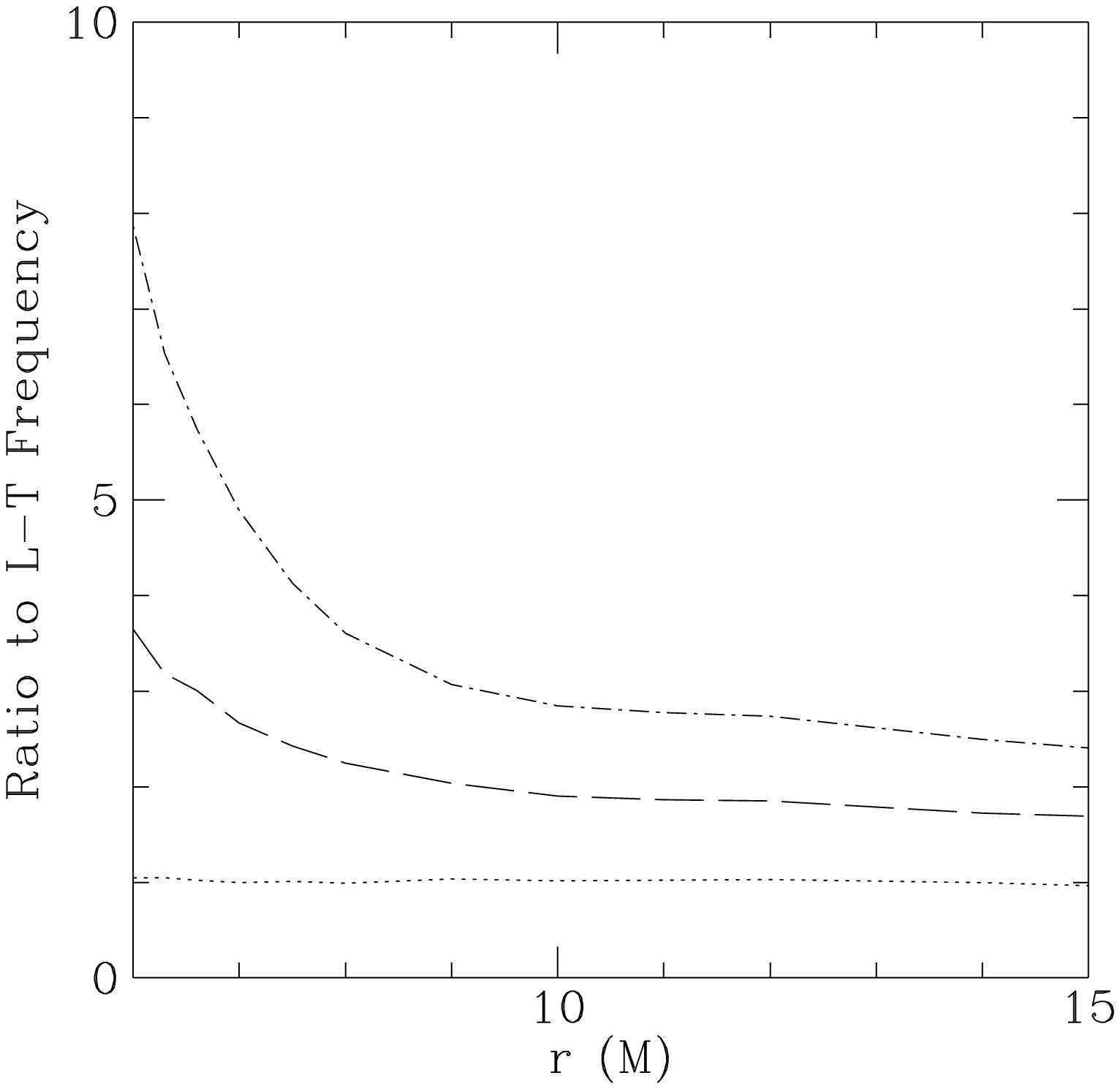,height=8.0truein,width=8.0truein}}
\end{figure*}

\begin{figure*}[t]
\hbox{\hskip -0.5truein
\psfig{file=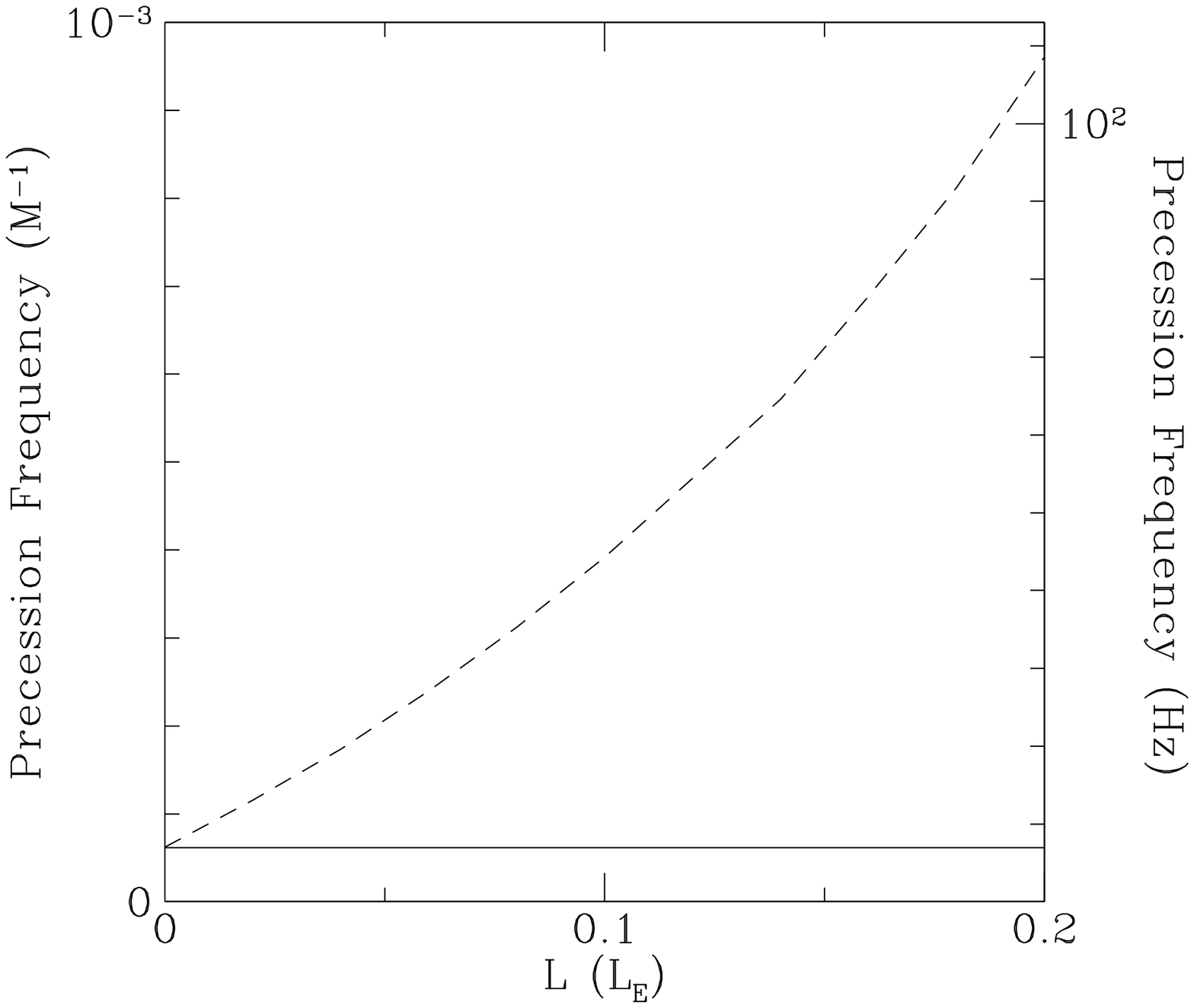,height=8.0truein,width=8.0truein}}
\end{figure*}

\end{document}